\documentclass[twocol]{arXiv}
\usepackage{sidecap}
\sidecaptionvpos{figure}{t}
\renewcommand{\thefigure}{{\bf \arabic{figure}}}
\renewcommand{\thetable}{{\bf \arabic{table}}}

\usepackage[
    papersize={8.5in,11in},
    twoside=false,
    inner=1.7cm, outer=1.7cm,
    top=2.5cm, bottom=1cm,
    footskip=1cm,
    nomarginpar,
]{geometry}

\usepackage{graphicx}%
\usepackage{amsmath,amssymb,amsfonts}%
\usepackage[separate-uncertainty = true, multi-part-units=single]{siunitx}%

\title{Spatial Controls of Lower Tropospheric Stability}

\authors{
    Senne Van Loon$^\ast$\correspondingauthor{
        Senne Van Loon; senne.van\_loon@colostate.edu
    } and 
    Maria Rugenstein}

\affiliation{
    Department of Atmospheric Science, 
    Colorado State University, Fort Collins, Colorado, USA}

\abstract{%
Marine low clouds play a crucial role in Earth's radiation budget. These clouds efficiently reflect sunlight and drive the magnitude and sign of the global cloud feedback. Nevertheless, the evolution of shallow cloud decks over the last decades is not well understood. A dominant control of this low cloud cover is the lower tropospheric stability, quantified by the estimated inversion strength (EIS). We quantify how regional EIS depends on local and remote surface temperature, revealing the dynamics controlling the shallow cloud characteristics on annual timescales. We find that global EIS increases with warming in tropical regions of ascent and decreases with warming in regions of descent. In addition to the West Pacific Warm Pool, the Atlantic convection regions and the central Pacific are important predictors. Focusing on subtropical ocean upwelling regions in different ocean basins, where the low cloud decks reside, EIS increases with a fairly complex pattern of remote warming and decreases with local warming. The spatial relationship between surface temperature and EIS is robust across climate models and reanalyses, allowing us to constrain the spread in historical EIS trend estimates. In the Southeast Pacific, historical surface temperature decreased, but we attribute the observed EIS increase since 1980 entirely to remote warming. Our results challenge the canonical dominance of the West Pacific Warm Pool in controlling low cloud feedbacks in the eastern Pacific and give mechanistic insights into the spatial dependence of radiative feedbacks on surface temperature patterns. %
\\\\%
\textnormal{\textit{Plain Language Summary:} 
Shallow clouds over subtropical oceans cool the Earth by reflecting sunlight, but how they respond to warming remains uncertain even after decades of work. These clouds are held in place by a temperature inversion, with relatively cool air at the surface and warmer air aloft. We investigate how this inversion, and thus the stability and extent of these shallow clouds, is influenced by local and remote surface temperatures. We quantify the spatial pattern of surface warming that governs the inversion strength, attribute observed changes in stability, and clarify which large-scale mechanisms contribute to the inversion strength. We now understand what sets past trends of stability, which helps to constrain future evolution of shallow clouds and hence, global temperature.%
\\\\%
\textit{Key points:}\\
$\bullet$\kern0.5em Subtropical lower tropospheric stability is influenced by surface warming patterns via both tropical and extratropical pathways\\
$\bullet$\kern0.5em Sensitivity of the lower tropospheric inversion strength to surface temperature is similar in climate models and reanalyses\\
$\bullet$\kern0.5em Local cooling did not significantly contribute to the observed Southeast Pacific lower tropospheric stability trend in 1980-2024
}}

\begin{document}

\maketitle

\section{Introduction}

Marine low clouds have the most efficient radiative effect of all cloud types, making them a crucial ingredient in characterizing climate sensitivity, performing climate projections, and understanding the hot-model problem \citep[e.g.,][]{Randall84, Sherwood20, Zelinka20, ipccAR6chap7, Rugenstein23c}. Marine stratiform cloud decks are often invoked to explain the pattern effect on radiative feedbacks, according to which the global top of the atmosphere radiation depends on the spatial pattern of surface warming \citep{Armour13, Andrews15, Stevens16, Mauritsen16, Zhou16, Zhou17, Dong19, Andrews22, Rugenstein23b}. Yet, our limited understanding of the dynamical controls of marine low clouds reflects in a large spread of low cloud feedback across climate models \citep{Scott20, Myers21, Myers23, Becker20, Sherwood20} and simulating them somewhat correctly requires a vertical resolution of the order of meters, hence, even storm resolving climate models struggle to represent them \citep{Nowak25}.

Stratiform clouds typically form in regions of oceanic upwelling. There, cold surface conditions beneath warmer free tropospheric air establish a steep temperature inversion in the lower troposphere. The strong inversion inhibits entrainment of free tropospheric air and causes a shallow, well-mixed atmospheric boundary layer. Moisture from the underlying ocean and advection fuel the formation of a thin layer of stratiform clouds at the top of the inversion \citep{Wood04,Wood06b,Wood12}. The shallow cloud decks are mainly controlled by local sea surface temperature and the strength of the lower tropospheric inversion, but relative humidity, vertical velocity, horizontal advection of surface temperature, and surface wind speed also play a role.

The strength of the lower tropospheric inversion is often approximated by the estimated inversion strength \citep[EIS,][]{Klein93, Wood06, Stevens07, Wood12, Zelinka20}. Recent work has highlighted the relevance of EIS in the time evolution of shallow clouds, the global energy budget, and the pattern effect \citep{Ceppi17, Ceppi19, Myers23, Kawaguchi25}. Increasing global EIS has caused a sign change of the near-global shortwave cloud radiative effect from positive to negative in the second half of the twentieth century, which has kept global mean temperature rise relatively low \citep{Myers23}. Despite the relevance of EIS for low cloud cover, it has been, until now, unquantified how surface warming across the globe controls EIS in regions with extensive marine low clouds, whether climate models simulate these controls correctly, and how EIS has evolved \citep{Qu15}.

Our goal is to understand the controls of the cloud controlling factor EIS. We evaluate the controls of EIS on global and local scales, by quantifying the spatial dependence of EIS on surface temperature in a novel statistical framework, using ridge regression and convolutional neural networks. This allows us to attribute observed EIS trends to local and remote surface temperature changes and give insights into dynamical drivers of low clouds and problems of their representation in climate models. Understanding the controls on EIS and its trends is an important step towards understanding the changes in the radiative effect of low clouds and hence, the pattern effect, both in the recent past and near future.

\section{Materials and Methods}\label{methods}

\subsection{Estimated inversion strength}

The estimated inversion strength \citep[EIS,][]{Wood06} is defined as
\begin{equation}
    {\rm EIS} = \theta_{700} - \theta_{1000} - \Gamma_{850} (z_{700} - {\rm LCL}),
    \label{eq:EIS}
\end{equation}
with $\theta_{700}$ and $\theta_{1000}$ the potential temperature at $\SI{700}{hPa}$ and $\SI{1000}{hPa}$, respectively, $\Gamma_{850}$ the moist adiabatic lapse rate at $\SI{850}{hPa}$ (using $T_{850} = (T_{1000}+T_{700})/2$), $z_{700}$ the height of the $\SI{700}{hPa}$ pressure level, and LCL the lifting condensation level. The LCL is calculated from the temperature at $\SI{1000}{hPa}$ assuming a constant relative humidity of $\SI{80}{\percent}$. Using surface temperature instead of $T_{1000}$ does not significantly alter our results.

We calculate EIS over all ocean areas between \ang{60}S and \ang{60}N (SI~Fig.~S2 shows the climatology of EIS) using yearly averages of $T_{700}$ and $T_{1000}$ on each model's/reanalysis' native grid (see below). We then take an area-weighted average over all ocean grid boxes in regions of interest, defined as (see, e.g., Figs.~\ref{1}~and~\ref{2} for specific locations): near global (NG; \ang{50}S-\ang{50}N), Southeast Pacific (SEP; \ang{40}S-\ang{0}S, \ang{110}W-\ang{70}W), Northeast Pacific (NEP; \ang{15}N-\ang{40}N, \ang{150}W-\ang{110}W), and Southeast Atlantic (SEA; \ang{40}S-\ang{0}S, \ang{20}W-\ang{20}E). We choose \ang{50}S-\ang{50}N because high latitudes negatively affect the relationship between EIS and global mean radiation \citep{Ceppi19}. Note that the NEP region is smaller than the SEP and SEA. We denote the area-weighted average EIS as $\langle{\rm EIS}\rangle$ or $\langle{\rm EIS}\rangle_{\rm region}$, specifying the region. 

\subsection{General circulation model data}\label{methods_gcm}

We use monthly-mean output of air temperature (ta; $T_{1000}$ and $T_{700}$) and near-surface ($\SI{2}{m}$) temperature (tas; $T$) from the historical run of four large initial condition ensemble general circulation models (GCMs; see SI~Table~S2 for all GCMs used): CanESM5 \citep{CanESM5}, GFDL-SPEAR-MED \citep{GFDLSPEARMED}, MIROC6 \citep{MIROC6}, and MPI-ESM1.2-LR \citep{MPI12}. These GCMs differ in the implementation of cloud parametrizations and in their climate sensitivity, offering a diverse set of simulations with enough ensemble members. We compute yearly averages in the overlapping period 1921-2014 from all ensemble members of each model. We use yearly values instead of monthly to avoid seasonal effects and lead-lag relationships shorter than one year. EIS is calculated from yearly-averaged $T_{1000}$ and $T_{700}$ on the model's native grid and then spatially averaged. We bilinearly regrid $T$ with periodic boundary conditions to a common $\ang{2.8}\times\ang{2.8}$ grid (native to CanESM5, lowest resolution among the models). All variables ($\langle{\rm EIS}\rangle$ and $T$) are detrended by removing the ensemble mean within each model. We assume internal variability within this time period does not change substantially. 

We use future scenarios to compare $\langle{\rm EIS}\rangle$ trends to reanalyses (shown in Fig.~\ref{5}E and SI~Fig.~S10). We use the SSP2-4.5 and SSP5-8.5 scenarios from the same four models, apart from GFDL-SPEAR-MED, for which only the SSP5-8.5 scenario is available. We compute yearly averages for all available ensemble members in 2015-2024 and concatenate them with the historical runs to obtain $\langle{\rm EIS}\rangle$ in 1921-2024. The future scenarios are not used to train the regression models, but only to compare simulated trends in 1980-2024 to reanalysis products.

\subsection{Reanalysis and observational data}\label{methodsreanalyses}

Yearly averages in 1940-2024 are calculated from monthly-mean $T_{1000}$, $T_{700}$, and $T$ from ERA5 \citep{Hersbach20} and processed in a similar way as the GCM data: EIS is calculated on the ERA5 native grid and averaged, and $T$ is bilinearly regridded to $\ang{2.8}\times\ang{2.8}$. ERA5 assimilates a combination of conventional and satellite (since 1979) observations and is expected to be a good approximation of the historical climate.

Yearly values of $\langle{\rm EIS}\rangle$ and $T$ are detrended with a high-pass filter. A Butterworth filter of order 4 with a cutoff period of 10 years is applied forwards and backwards using SciPy's signal package \citep{SciPy}. Then, the first and last four years are removed to avoid edge effects. For $T$, this is done at each grid box separately, after regridding. We assume that the filtered data describes internal variability only, but note that this procedure removes all low-frequency variability, whether it is forced or internal. 

To understand observed trends in $\langle{\rm EIS}\rangle$ (section~\ref{sec:trends}), we compare to two additional reanalysis products: JRA-3Q \citep{Kosaka24} and MERRA-2 \citep{Gelaro2017}. Both datasets are processed in the same way as ERA5 data, but not filtered, to obtain $\langle{\rm EIS}\rangle$ and $T$ from JRA-3Q and MERRA-2. 

Finally, we use three gridded sea surface temperature (SST) datasets to calculate $\langle{\rm EIS}\rangle$ predicted by our regression models (see Section~\ref{sec:ridge} and \ref{sec:app}): COBE2 \citep{COBE2}, NOAAGlobalTemp \citep{NOAAGlobalTemp}, and HadISST \citep{Rayner03}. All SST datasets are bilinearly regridded to $\ang{2.8}\times\ang{2.8}$ and not filtered.

\subsection{Linear ridge regression}\label{sec:ridge}

We perform multiple linear regression to determine the sensitivity of $\langle{\rm EIS}\rangle_{\rm region}$ to surface temperature $T$ (\ref{sec:app} and SI~Fig.~S1 for more details). The regression model predicts $\langle{\rm EIS}\rangle_{\rm region}$ from $T_j$ at each grid box $j$ as $\langle{\rm EIS}\rangle_{\rm region} \simeq \sum_j a_j T_j + b$. Here, $\langle{\rm EIS}\rangle$ is a single value, $T$ is a map of surface temperature over ocean areas between \ang{60}S and \ang{60}N, and $j$ indicates the grid box location. Because ordinary linear regression is prone to overfitting, we use a ridge parameter to regularize. Ridge regression adds a penalty term to the least squares loss function that forces the regression coefficients to remain small and reduces overfitting. 

\begin{SCfigure*}
    \centering
    \includegraphics{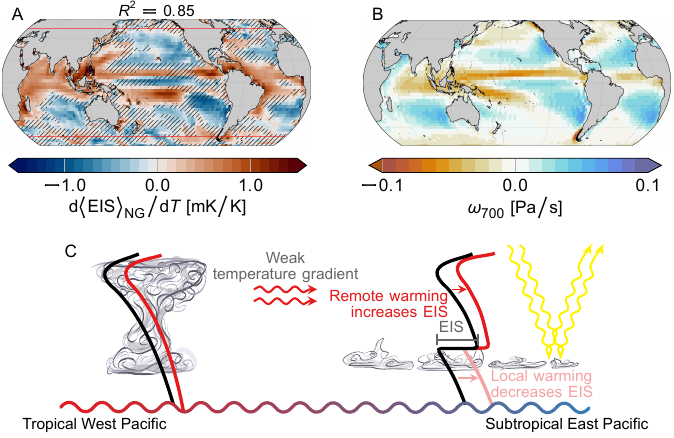}
    \caption{{\bf Controls of estimated inversion strength (EIS). }
    (A) Sensitivity of near-global mean EIS ($\langle {\rm EIS} \rangle_{\rm NG}$, \ang{50}S-\ang{50}N, indicated by red lines) to surface temperature. Red regions (positive sensitivity) indicate where increasing surface temperature leads to an increasing $\langle {\rm EIS} \rangle_{\rm NG}$, and vice versa for blue regions (negative sensitivity). Hatching indicates regions where sensitivity maps obtained from the four models separately do not agree on the sign.
    (B) Multi-model-mean annual climatology of vertical velocity at $\SI{700}{hPa}$ ($\omega_{700}$) in 1991-2014, showing regions of ascent (red) and descent (blue). 
    (C) Schematic of how the weak temperature gradient (mediated by tropical gravity waves) sets the free tropospheric temperature across the Pacific, as often evoked to explain the pattern effect. 
    The black line depicts the climatological mean temperature structure of the atmosphere. EIS increases with increasing free tropospheric temperature through warming in remote regions of deep convection (red) or by locally cooling the surface (not illustrated). Conversely, EIS decreases when cooling the free troposphere (not illustrated) or warming the surface (pink). Later, we argue that the dynamics setting free tropospheric temperature in the subtropics are more complex and involve planetary Rossby wave dynamics (Fig.~\ref{4}).
    }
    \label{1}
\end{SCfigure*}

The resulting regression coefficients $a_j$ can be interpreted as a ``sensitivity map'' $a_j \sim {\rm d}\langle{\rm EIS}\rangle_{\rm region}/{\rm d}T_j$: they indicate the change in $\langle{\rm EIS}\rangle_{\rm region}$ per degree of warming in each grid box $j$. These are similar to sensitivity maps for radiation or precipitation recently popularized using Green's functions \citep[e.g.,][]{Zhou17,Dong19,Alessi23,Bloch-Johnson24,Alessi24,Fredericks26}, but statistical. Green's functions use causal relationships occurring in atmospheric climate model simulations, but do not represent coupled processes (since they are computed with fixed SST) and the prescribed SST-patch perturbations never occur in reality. In contrast, our regression model predicts $\langle{\rm EIS}\rangle_{\rm region}$ from $T$ based on correlations in the data, and therefore have no causal structure. Thus, our sensitivity maps are entirely statistical and contain noise that is hard to quantify, but have the advantage of sampling the coupled system's actually occurring, coherent modes of variability, such as El Ni\~{n}o Southern oscillation (ENSO) and Pacific decadal oscillation [further discussed in \cite{Rugenstein25}]. Advantages and disadvantages of both methods, including other statistical methods, are discussed by \cite{Fredericks26} in the context of global radiation. 

Once the regression models are trained, we can apply it to any surface temperature map to predict $\langle{\rm EIS}\rangle_{\rm region}$. This allows us to predict $\langle{\rm EIS}\rangle_{\rm region}$ from gridded surface temperature products based on observations. We refer to ``predicted'' values as $\langle{\rm EIS}\rangle_{\rm region}$ or $\langle{\rm EIS}\rangle_{\rm region}$ trends calculated by applying the regression model to different surface temperature maps.

\section{Sensitivity of EIS to near-surface temperature}

We evaluate the internal variability (IV) relationship between spatial maps of near-surface temperature ($T$) and area-weighted spatial averages of EIS in different regions ($\langle{\rm EIS}\rangle_{\rm region}$), using ridge regression (section~\ref{sec:ridge}) on annual timescales. We first train the regression model on large initial condition ensembles of four coupled general circulation models (GCMs, section~\ref{methods_gcm}), and then in an observation-based reanalysis product. For the GCM data, we subtract the single-model ensemble mean from each member to include IV only (section~\ref{methods_gcm}). For reanalyses, we estimate the IV-contribution as the 10-year high-pass filtered data (section~\ref{methodsreanalyses}). We use initial condition ensembles because they provide enough data to robustly train a regression model. We focus on regions with extensive climatological marine low clouds, but start by evaluating the near-global EIS.

\subsection{Near-global EIS}

Near-global EIS has been shown to be a first-order control on global mean feedbacks. When added to a standard energy balance model, it can explain variations in the global feedback parameter and climate sensitivity \citep{Ceppi19}. We investigate the spatial controls of near-global EIS ($\langle{\rm EIS}\rangle_{\rm NG}$, averaged over ocean areas between \ang{50}S-\ang{50}N) by regressing it onto near-surface temperature maps in four GCMs, separately and simultaneously.

The sensitivity of $\langle{\rm EIS}\rangle_{\rm NG}$ to spatial surface temperature follows our expectations (Fig.~\ref{1}A). Warming in tropical regions of ascent (red in Fig.~\ref{1}B) increases $\langle{\rm EIS}\rangle_{\rm NG}$, while warming in descending areas (blue in Fig.~\ref{1}B) decreases $\langle{\rm EIS}\rangle_{\rm NG}$, especially in regions of high stability. The entire Indian ocean contributes positively to $\langle{\rm EIS}\rangle_{\rm NG}$, even though it is less convectively active than the West Pacific Warm Pool (Fig.~\ref{1}B, using large vertical velocity as a proxy for convection). Warming in the subtropical central and eastern part of the ocean basins, especially the Pacific, decreases $\langle{\rm EIS}\rangle_{\rm NG}$. The total sensitivity of $\langle{\rm EIS}\rangle_{\rm NG}$ to surface temperature (i.e., the sum of the sensitivity map over all grid boxes) is $\SI{0.11}{K/K}$, indicating that a uniform warming of $\SI{1}{K}$ leads to an increase in $\langle{\rm EIS}\rangle_{\rm NG}$ of $\SI{0.11}{K}$, in agreement with \cite{Ceppi19}. 

\subsection{EIS in regions of climatological marine low clouds}

\begin{figure*}
    \centering
    \includegraphics[width=\textwidth]{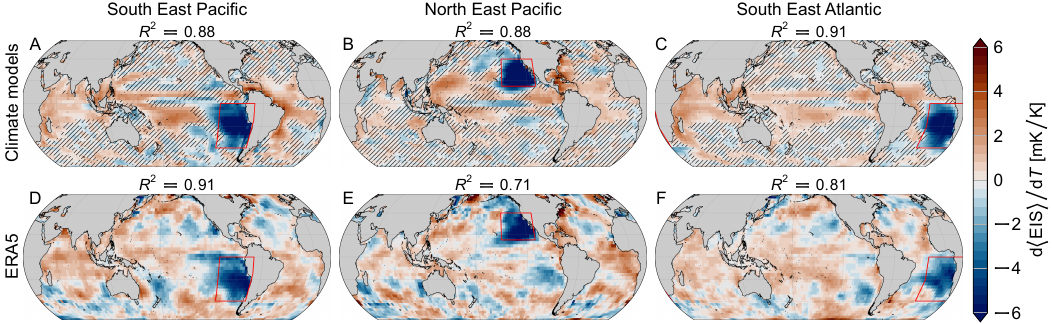}
    \caption{{\bf Controls of regional estimated inversion strength. }
    Top row shows the annual sensitivity of $\langle {\rm EIS} \rangle_{\rm region}$ to surface temperature in regions indicated by red boxes: (A) Southeast Pacific, (B) Northeast Pacific, (C) Southeast Atlantic, based on ridge regression on data from four climate models. Hatching indicates regions where sensitivity maps obtained from the four models separately do not agree on the sign. 
    Bottom row (D-F) shows the sensitivity maps in the same regions, but using 10-year high-pass filtered ERA5 data. $R^2$ values are calculated for held-back testing members and displayed above each map (see~\ref{sec:app}).}
    \label{2}
\end{figure*}

The Southeast Pacific (SEP), Northeast Pacific (NEP), and Southeast Atlantic (SEA) dominate $\langle{\rm EIS}\rangle_{\rm NG}$ because they have high climatological EIS (SI~Fig.~S2 shows EIS climatology in 1991-2014 in models and reanalyses). Locally, EIS is set by surface and free tropospheric temperature. The latter is thought to be influenced by regions of deep convection in the tropics, and mediated by the dynamics of the weak temperature gradient approximation \citep[WTGA,][]{Charney63,Sobel01}. 

The WTGA states that temperature gradients in the tropical free troposphere are flattened quickly by gravity waves, because the Coriolis force is small near the equator. Horizontal gradients of temperature aloft are thus difficult to maintain and substantially smaller than at the surface. Surface temperature anomalies are efficiently transported vertically through updrafts in deep convective clouds (Fig.~\ref{1}C). 

The WTGA is a central explanation in the debate around the pattern effect, explaining nonlocal radiative feedbacks \citep[e.g.,][]{Mauritsen16, Zhou17, Dong19, ipccAR6chap7, Andrews22}. It is qualitatively evoked to explain how warming in the West Pacific Warm Pool influences remote regions, especially the subtropical eastern Pacific (Fig.~\ref{1}C). However, the WTGA might not be sufficient to explain the pattern effect. Patch simulations in an atmosphere-only model, where the surface temperature is locally increased in the equatorial Pacific, indicate that moist static energy does not spread uniformly through the tropical free troposphere \citep{Williams23}. The same simulations show that free tropospheric stability can decrease in some tropical subsidence regions with surface warming in convective regions, counter to the WTGA \citep{Mackie25}. 

We evaluate the relevance of the tropical gravity waves in setting local EIS by regressing $\langle{\rm EIS}\rangle_{\rm region}$ in the SEP, NEP, and SEA, to near-surface temperature maps in four GCMs. This allows us, for the first time, to quantify controls of EIS on local scales. We target regions that are most important in setting low cloud feedbacks, with climatologically high EIS and extensive low cloud decks (see SI~Fig.~S2 for the EIS climatology). Fig.~\ref{2}A-C shows the sensitivity of $\langle{\rm EIS}\rangle$ averaged over boxes (indicated in red) in the SEP, NEP, and SEA to surface temperature. Focusing on regions of climatological low cloud decks results in a more complex spatial dependence than the global sensitivity map (Fig.~\ref{1}A). \cite{Ding25} found no significant connection between the climatological subtropical highs west of stratocumulus regions and EIS, but our sensitivity maps suggest that multiple regions in the extratopics influence $\langle{\rm EIS}\rangle$. Models agree on key regions and magnitudes of this spatial sensitivity (see hatching). The sensitivity maps confirm previous findings that remote temperature anomalies can control subtropical $\langle{\rm EIS}\rangle$ \citep[e.g.,][]{Mauritsen16, Zhou17, Dong19} and quantify the spatial pattern of these controls.

Locally, a surface temperature increase within the region contributes negatively to $\langle{\rm EIS}\rangle_{\rm region}$ (blue in Fig.~\ref{2}A-C, pink line in Fig.~\ref{1}C). This local sensitivity is the same order of magnitude in all three locations. A uniform local warming of $\SI{1}{K}$ (in the red boxes in Fig.~\ref{2}A-C) decreases $\langle{\rm EIS}\rangle_{\rm region}$ by $\SI{-0.81}{K}$, $\SI{-0.82}{K}$, and $\SI{-0.76}{K}$ in the SEP, NEP, and SEA, respectively. This means that local surface warming also warms the free troposphere (see SI~Fig.~S3 for the sensitivity of free tropospheric temperature $T_{700}$ to surface temperature). Using Eq.~(\ref{eq:EIS}), an increase of surface temperature by $\SI{1}{K}$ while keeping the free troposphere fixed decreases EIS by $\sim\SI{-1.2}{K}$ \citep[see also][]{Qu15}. Since $\langle{\rm EIS}\rangle_{\rm region}$ only decreases by $\sim\SI{-0.8}{K}$, the local free troposphere warms by $\sim\SI{0.4}{K}$ with a $\SI{1}{K}$ surface warming.

Remotely, a surface temperature increase contributes to an increase in $\langle{\rm EIS}\rangle_{\rm region}$ (red in Fig.~\ref{2}A-C, red line in Fig.~\ref{1}C). Notably, the remote sensitivity of $\langle{\rm EIS}\rangle_{\rm region}$ in these locations is not the same as the one for $\langle{\rm EIS}\rangle_{\rm NG}$ (Fig.~\ref{1}A) and does not exclusively indicate regions of climatological deep convection. For example, the most efficient remote control of $\langle{\rm EIS}\rangle_{\rm SEP}$ is not the center of the West Pacific Warm Pool, but regions closer to the SEP, which also exhibit convection (Fig.~\ref{1}B). Similarly, the West Pacific Warm Pool gets picked up robustly across all models to predict $\langle{\rm EIS}\rangle_{\rm NEP}$, but sub- and extratropical surface temperature patterns are as relevant to $\langle{\rm EIS}\rangle_{\rm NEP}$ as the tropical regions of deep convection. A uniform remote warming of $\SI{1}{K}$ (everywhere but the red boxes in Fig.~\ref{2}A-C) increases $\langle{\rm EIS}\rangle_{\rm region}$ by $\SI{0.94}{K}$, $\SI{0.69}{K}$, and $\SI{0.90}{K}$ in the SEP, NEP, and SEA, respectively. This means a remote warming is very efficient at changing the free tropospheric temperature in the regions of climatological low cloud decks. Keeping the local surface temperature fixed, $T_{700}$ must increase by $\sim\SI{0.92}{K}$ to increase EIS by $\SI{0.90}{K}$, suggesting that the $\SI{1}{K}$ remote surface warming is efficiently transported to the local free troposphere. In the SEP and SEA, the remote contribution is larger than the local contribution, resulting in an increased stability with global uniform warming, while the NEP is dominated by local effects (see SI~Table~S1 for a summary of local and remote contributions).

The sensitivity maps from regression coefficients (Fig.~\ref{2}A-C) explain correlations in the data, and do not imply any causation. Within a given year, a positive regression coefficient indicates that a local temperature anomaly tends to coincide with higher $\langle{\rm EIS}\rangle_{\rm region}$ in the same year. These correlations can be caused by many processes, both in the atmosphere \citep[e.g., tropical gravity waves;][]{Mauritsen16, Zhou17, Dong19} and at the surface \citep[e.g., ocean circulation;][]{Kang23}. The training data contains such coupled processes, but the regression model does not have any information on the underlying dynamics. In contrast, Green's function simulations can be used to investigate causal relationships, but only capture atmospheric processes, since the ocean is prescribed. Sensitivity maps from Green's functions (see SI~Fig.~S6) show similar large-scale features as the regression-based sensitivity maps. This indicates that the regression model is not picking up spurious correlations, but physical relationships between surface temperature and $\langle{\rm EIS}\rangle_{\rm region}$. 

The local versus remote contributions to $\langle{\rm EIS}\rangle$ become clearer when examining the spatial controls of $T_{1000}$ and $T_{700}$ separately (see SI~Figs.~3-4 for the sensitivity maps). The local negative contribution is traceable to $T_{1000}$ (warming beneath the inversion, pink line in Fig.~\ref{1}C) and the remote positive contribution to $T_{700}$ (warming above the inversion, red line in Fig.~\ref{1}C), confirming our physical interpretation of the statistical sensitivity maps.

\subsection{Robust sensitivity across models and reanalyses}

GCMs heavily parameterize deep and shallow convection and do not necessarily simulate the correct variability of surface temperature variations, e.g., due to ENSO \citep[e.g.,][]{Hourdin17,Deser20,Maher23}. The results discussed so far are robust across the four GCMs used here, even though their cloud parameterizations and representation of internal variability (IV) differ substantially \citep[e.g.,][]{Maher23,Zelinka20}. This agreement raises hope but does not imply that the observed spatial controls of EIS are the same as in GCMs. We turn to reanalysis data to evaluate the robustness of our results. 

We again perform ridge regression, but use ERA5 data in 1940-2024 to find the sensitivity of $\langle{\rm EIS}\rangle_{\rm region}$ to surface temperature (Fig.~\ref{2}D-F). We first detrend the data using a 10-year high-pass filter (section~\ref{methodsreanalyses}), to compare the observed IV with the IV of GCMs investigated in the previous section. Using a reanalysis has the benefit that we can probe the controls on EIS in a model product including observations, but we can only use 85 years versus $\sim10,000$ years for GCMs. Therefore, the reanalysis-based sensitivity maps (Fig.~\ref{2}D-F) probably contain more noise than the GCM-based maps (Fig.~\ref{2}A-C). 

The large-scale features and magnitudes of the sensitivities ${\rm d}\langle{\rm EIS}\rangle/{\rm d}T$ found in GCMs also appear in the reanalysis-based estimates. Local surface warming decreases $\langle{\rm EIS}\rangle$, while remote warming increases it. The most sensitive remote locations appear mainly in the tropics and regions of deep convection, but are more spread out than expected from a strict interpretation of the WTGA. The Atlantic and Indian Ocean and sub- and extra-tropics are as relevant to predict variations in $\langle{\rm EIS}\rangle_{\rm SEP}$ as the West Pacific Warm Pool. Because of limited data, we cannot evaluate the reliability of all remote contributions to $\langle{\rm EIS}\rangle$, and expect the sensitivity maps to change with a longer observational record. However, the large-scale features of the sensitivity maps are robust across GCMs and ERA5. GCMs' sensitivities to remote uniform warming generally agree with ERA5, but their local contribution is stronger (SI~Table~S1 summarizes the separate contributions). This results in a higher sensitivity to global uniform warming in the reanalysis-based regression than the GCM-based regression, the former being positive for all regions.

\begin{figure}
    \centering
    \includegraphics{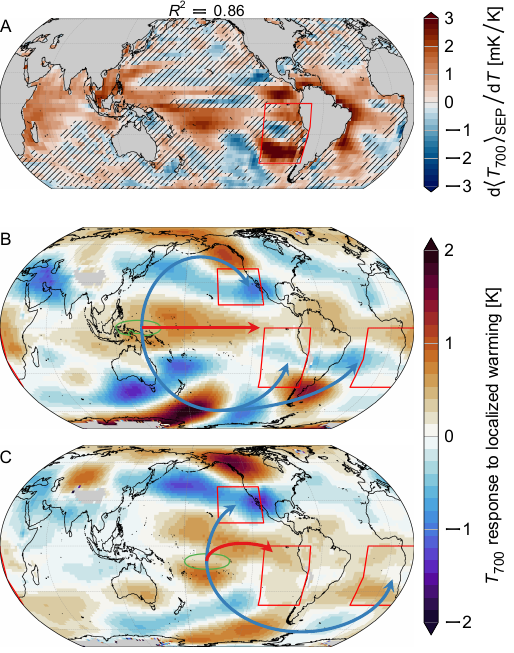}
    \caption{{\bf Free tropospheric temperature response to surface warming.}
    (A) Sensitivity of $\langle T_{700}\rangle_{\rm SEP}$ to surface temperature based on ridge regression on data from four climate models. Hatching indicates regions where sensitivity maps obtained from the four models separately do not agree on the sign. 
    (B and C) Average response of $T_{700}$ to localized surface warming in the West Pacific Warm Pool and central Pacific, indicated by green ovals. Red boxes show the SEP, NEP, and SEA regions. Arrows are for illustration only, to indicate possible pathways for surface warming to affect $T_{700}$ in low cloud deck regions via fast tropical gravity waves (red) or Rossby waves (blue).
    }
    \label{3}
\end{figure}

\begin{SCfigure*}
    \centering
    \includegraphics[width=12cm]{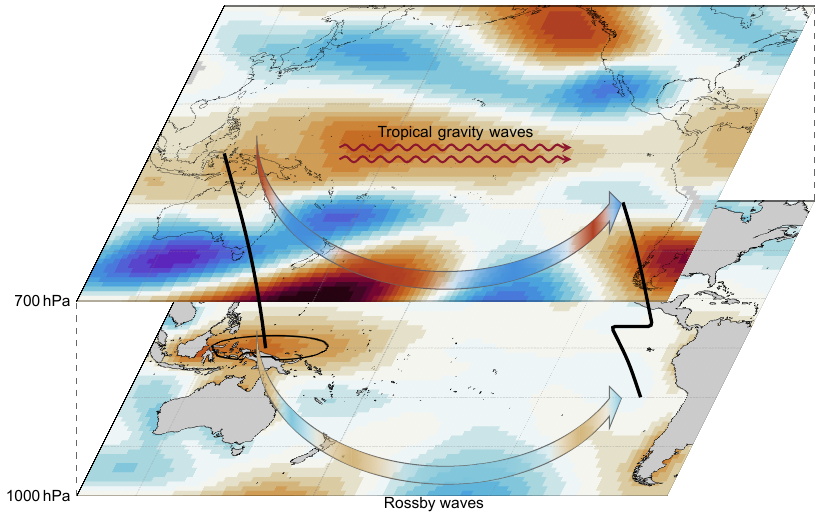}
    \caption{{\bf Schematic of proposed remote atmospheric teleconnections affecting Southeast Pacific EIS.}
        Maps qualitatively illustrate the response of $T_{1000}$ and $T_{700}$ to surface warming in the West Pacific Warm Pool. Black lines sketch the climatological temperature profiles in the West Pacific and Southeast Pacific. Warm pool surface warming can affect the free tropospheric temperature in the Southeast Pacific through fast tropical gravity waves (wavy red arrows) and Rossby waves (alternating red-blue arrows). Additionally, other processes (e.g., oceanic pathways or land-sea contrast) can change $\langle{\rm EIS}\rangle$. Similar dynamics affect $\langle{\rm EIS}\rangle$ in other regions of climatological marine low clouds.
    }
    \label{4}
\end{SCfigure*}

Our analysis is performed on annual data, which removes the variability on seasonal timescales, even though the latter could be large \citep[e.g.,][]{Wood12}. In ERA5, the variance remaining after taking annual means from monthly data depends on the region ($\SI{55}{\percent}$ for SEP, $\SI{42}{\percent}$ for NEP, and $\SI{35}{\percent}$ for SEA). Applying the regression model trained on annual data to monthly ERA5 surface temperature data results in similar $R^2$ values for the predicted $\langle{\rm EIS}\rangle_{\rm region,\,month}$ ($R^2\simeq\SI{57}{\percent}$ for SEP, $R^2\simeq\SI{47}{\percent}$ for NEP, and $R^2\simeq\SI{49}{\percent}$ for SEA). Training the regression model on monthly data results in similar sensitivity maps as training on annual data, but with more noise (see SI~Fig.~S5). This indicates that there is not much additional information in the monthly variability of surface temperature for predicting $\langle{\rm EIS}\rangle_{\rm region}$. However, correctly representing sub-annual variability should take into account possible lead-lag relationships between surface temperature and $\langle{\rm EIS}\rangle_{\rm region}$, which is beyond the scope of this paper.

\section{Rossby waves likely modulate subtropical EIS}

The sensitivity maps ${\rm d}\langle{\rm EIS}\rangle/{\rm d}T$ point to more complex dynamics than tropical gravity waves alone. We argue that Rossby waves responding to tropical surface warming are an essential ingredient to understanding controls of EIS in addition to tropical gravity waves.

The sensitivities of $\langle{\rm EIS}\rangle_{\rm SEP}$ and $\langle{\rm EIS}\rangle_{\rm NEP}$ to surface temperature (Fig.~\ref{2}A,B~and~D,E) indicate that tropical-to-extratropical teleconnections originating in the West Pacific Warm Pool may play a role in modulating the stability in the SEP. This wave train appears as an alternating pattern of positive and negative sensitivities. The alternating pattern is more obvious in the sensitivity of the free tropospheric temperature to surface temperature (Fig.~\ref{3}A, see SI~Fig.~S3 for other regions), obtained by regressing $\langle T_{700}\rangle_{\rm SEP}$ to $T$. All else equal, locally increasing $\langle T_{700}\rangle$ increases EIS and vice versa [see Fig.~\ref{1}C and Eq.~(\ref{eq:EIS})]. In the SEP, the wave pattern changes sign within the SEP box, with a negative sensitivity in the middle of the box. All models except for CanESM5 agree on this local negative sensitivity, as does ERA5. 

We qualitatively investigate the pathways of temperature anomalies in idealized atmosphere-only model simulations. Figs.~\ref{3}B-C show the average annual mean anomalous response of $T_{700}$ to a localized surface warming in the equatorial West Pacific and off-equatorial central Pacific (indicated by green contours) in ECHAM6 \citep{Stevens13,Alessi23,Bloch-Johnson24}. The maps represent the causal response to surface warming and reveal signatures of both the WTGA and a Rossby wave train. Other atmospheric models show similar responses \citep[see, e.g.,][]{Zhou17,Dong19,Andrews18}. Surface warming in the West Pacific Warm Pool (Fig.~\ref{3}B) warms the entire tropical free troposphere through fast gravity waves according to the WTGA, although not uniformly. The same warming also triggers a Rossby wave train that, on average, tends to cool the free troposphere in regions of climatological low clouds, destabilizing those regions. These competing effects of tropical gravity waves and extratropical Rossby waves reduce the West Pacific Warm Pool's efficiency in warming the free troposphere in the SEP, NEP, and SEA, which modulates $\langle{\rm EIS}\rangle_{\rm region}$. Surface warming in the off-equatorial central Pacific (Fig.~\ref{3}C) can efficiently warm the SEP and SEA through tropical gravity waves (i.e., the WTGA), explaining the large positive sensitivity of $\langle{\rm EIS}\rangle_{\rm SEP}$ and $\langle{\rm EIS}\rangle_{\rm SEA}$ to warming in the central Pacific (Fig.~\ref{2}A~and~C). In contrast, in the NEP, the Rossby wave response cools the free troposphere, weakening the sensitivity of $\langle{\rm EIS}\rangle_{\rm NEP}$ to warming in the central Pacific (Fig.~\ref{2}B).

In summary, we argue that the dynamic response to tropical warming matters for eastern basin EIS (Fig.~\ref{4}). The West Pacific Warm Pool appears less efficient in the sensitivity maps because the extratropical pathways (red-blue arrows in Fig.~\ref{4}) opposes the response expected from the WTGA alone (wavy arrows in Fig.~\ref{4}). We do not quantify the relative importance of the gravity waves and Rossby waves in setting $\langle{\rm EIS}\rangle_{\rm region}$ or the different timescales they operate on. Other processes, such as land-sea contrast or oceanic pathways, could contribute as well. Rather, we suggest that Rossby waves can modulate $\langle{\rm EIS}\rangle$ in regions of low cloud decks and should be considered in future work on the dynamics of the pattern effect and low cloud feedbacks \citep{Andrews18}.

\section{Attributing EIS trends in the Southeast Pacific}\label{sec:trends}

\begin{SCfigure*}
    \centering
    \includegraphics{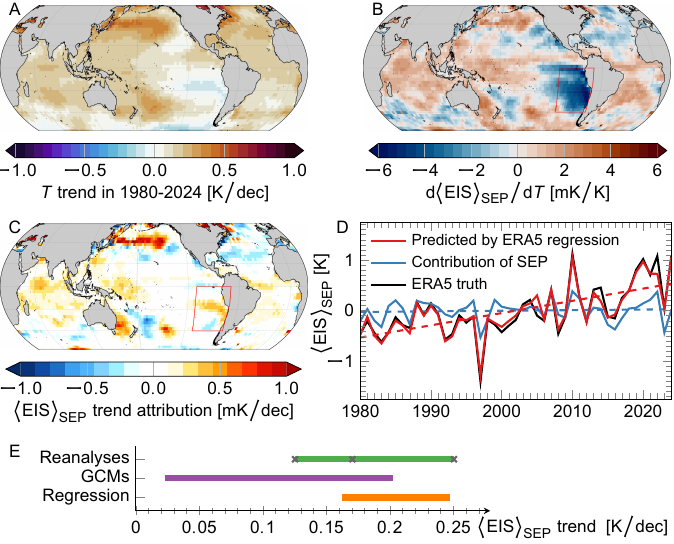}
    \caption{{\bf Attribution of the estimated inversion strength trend in 1980-2024 in the Southeast Pacific from ERA5. }
    (A) Surface temperature trends in 1980-2024 in ERA5.
    (B) Sensitivity of $\langle{\rm EIS}\rangle_{\rm SEP}$ (EIS averaged over the red box in the Southeast Pacific) to surface temperature (as Fig. 2D but using unfiltered training data)
    (C) Attribution map of $\langle{\rm EIS}\rangle_{\rm SEP}$ calculated by multiplying the observed surface temperature trend (A) with the sensitivity map (B). 
    (D) $\langle{\rm EIS}\rangle_{\rm SEP}$ in 1980-2024 in ERA5 (black), compared to the prediction from ridge regression using the global surface temperature map (red) versus the surface temperature in the SEP box only (blue; see text for details). All values are anomalies with respect to the 1980-2024 average.
    (E) Mean $\langle{\rm EIS}\rangle_{\rm SEP}$ trend (1980-2024) in reanalyses and GCMs. Green shows the spread among different reanalyses (crosses indicate individual reanalyses, from left to right: MERRA2, JRA-3Q, and ERA5). Purple shows the spread among different ensemble members in four different GCMs. Orange shows the spread among predicted trends by applying our regression models (trained either on ERA5 or GCMs) to different observed surface temperature datasets. See \ref{sec:app} and SI~Fig.~S10 for details. 
    }
    \label{5}
\end{SCfigure*}

The SEP is the region with the largest extent of climatological marine low clouds. This region has cooled at the surface since 1980, according to observations. However, GCMs do not simulate this cooling, not even in large initial condition ensembles, indicating that the observed cooling is likely not due to internal variability \citep[e.g.,][]{Wills22}. The nature of the discrepancy between models and observations is a topic of debate \citep[e.g.,][]{Olonscheck20,Watanabe21,Seager22,Wills22,Watanabe24}. The low cloud decks in the SEP are thought to matter for the equatorward propagation of extratropical surface temperature anomalies, but are simulated poorly \citep[e.g.,][]{Dong22,Kim22,Zheng25}. The mean state of the SEP low clouds influences the global low cloud feedbacks \citep{Ceppi24}, yet it is unknown to what degree these clouds are controlled by local versus remote conditions, and whether GCMs simulate these sensitivities correctly \citep{Myers21,Myers23}. 

We use various reanalyses to confirm a robust positive trend in $\langle{\rm EIS}\rangle_{\rm SEP}$ in 1980-2024, although the reanalyses disagree on its magnitude (Fig.~\ref{5}E, green, and SI~Fig.~S10, which shows the trend across reanalyses). Even though coupled historical GCM simulations are unable to reproduce the surface cooling in the SEP, they capture the increase in $\langle{\rm EIS}\rangle_{\rm SEP}$, but tend to underestimate the trend (Fig.~\ref{5}E, purple). EIS is the largest source of disagreement between reanalyses and models for low cloud feedbacks \citep{Myers23,Ceppi24,Kawaguchi25}, so we need to know what controlled the observed increase in $\langle{\rm EIS}\rangle_{\rm SEP}$ and why GCMs underestimate the trend.

To find out, we again perform ridge regression, but now on unfiltered ERA5 data from 1940-2024 (Fig.~\ref{5}B). That is, we now include variability longer than 10 years, while sensitivity maps in Fig.~\ref{2} were trained on 10-year high-pass filtered data. Although the resulting sensitivity maps are similar to the IV maps (see SI~Fig.~S5 for a comparison), we include long-term variability here to more accurately attribute the observed trend. Then, we pointwise multiply the sensitivity map ${\rm d}\langle{\rm EIS}\rangle_{\rm SEP}/{\rm d}T$ (Fig.~\ref{5}B) with the local trend of surface temperature in 1980-2024 (Fig.~\ref{5}A). The resulting map (Fig.~\ref{5}C) shows the contribution of local and remote temperature changes to the observed trend in $\langle{\rm EIS}\rangle_{\rm SEP}$ in 1980-2024 (Fig.~\ref{5}D, dashed red line). That is, summing over all ocean areas in Fig.~\ref{5}C results in the coefficient of the dashed red trend line in Fig.~\ref{5}D. Similarly, multiplying the spatial temperature anomaly in each year with the sensitivity map and summing over all ocean areas results in the predicted timeseries from the regression model (Fig.~\ref{5}D, red line). The trend in the predicted $\langle{\rm EIS}\rangle_{\rm SEP}$ is significantly positive (Table~\ref{tab1}) and the predicted timeseries closely matches the true ERA5 data (Fig.~\ref{5}D, black line). 

The steady decadal-long positive trend could, in principle, be due to local SEP cooling, remote warming, or both. Summing the attribution map over only the SEP region (red box in Fig.~\ref{5}B) results in a nonsignificant trend of $\SI{0.01\pm0.04}{K/decade}$ (Fig.~\ref{5}D, dashed blue line, and Table~\ref{tab1}), indicating that the observed trend in $\langle{\rm EIS}\rangle_{\rm SEP}$ is dominated by remote warming. Our results show that the cooling of the SEP has a minimal effect on the observed increase in $\langle{\rm EIS}\rangle_{\rm SEP}$, because remote locations warmed more consistently than the SEP cooled and contribute a much larger area (see SI~Table~S1 for an overview). Still, the attribution map (Fig.~\ref{5}C) shows that local cooling in the SEP region contributes positively to the observed increase in $\langle{\rm EIS}\rangle_{\rm SEP}$, but only weakly. 

\begin{SCtable*}\setlength{\parskip}{0pt}
    \centering
    \caption{Trends in estimated inversion strength (EIS) over the Southeast Pacific (SEP), Northeast Pacific (NEP), and Southeast Atlantic (SEA) in 1980-2024. Predicted trends are from the regression model trained on ERA5 applied to ERA5 surface temperature and separated into the contribution from local and remote temperature changes. All values are in $\si{K/decade}$ with $\SI{5}{\percent}$-$\SI{95}{\percent}$ confidence bounds, bold indicates trends that are significantly different from zero.}
    \begin{tabular}{r| c c | c c}
        Location & ERA5 trend & Predicted trend & Local contribution & Remote contribution \\\hline
        SEP & ${\bf 0.25\pm0.08}$ & ${\bf 0.24\pm0.07}$ & $      0.01\pm0.04 $ & ${\bf  0.23\pm0.05}$ \\
        NEP & ${\bf 0.15\pm0.08}$ & ${\bf 0.15\pm0.08}$ & ${\bf -0.05\pm0.04}$ & ${\bf  0.20\pm0.04}$ \\
        SEA & $     0.05\pm0.05 $ & ${\bf 0.06\pm0.05}$ & ${\bf -0.06\pm0.02}$ & ${\bf  0.12\pm0.04}$ \\
    \end{tabular}
    \label{tab1}
\end{SCtable*}

\subsection{Constraining historical EIS trends}

Historical EIS can be reproduced by the linear regression model trained on internal variability, both in ERA5 and in GCMs. The sensitivity map based on IV (Fig.~\ref{2}A~and~D) can predict EIS from ERA5 when applied to unfiltered $T$ from ERA5 (see SI~Fig.~S7 for the prediction from different regression models). Also the trend in $\langle{\rm EIS}\rangle_{\rm SEP}$ is captured by the regression models trained on IV. All three regression models (trained on unfiltered ERA5 data, detrended ERA5 data, or detrended GCM data) predict a robust trend in $\langle{\rm EIS}\rangle_{\rm SEP}$, even when applying the regression models to different surface temperature datasets. This is also found when using a Green's function sensitivity map to predict $\langle{\rm EIS}\rangle_{\rm SEP}$. Applying our regression models to historical surface temperature constrains the spread of the $\langle{\rm EIS}\rangle_{\rm SEP}$ trend to the higher end of reanalyses predictions (Fig.~\ref{5}E, orange, and SI~Fig.~S10). This indicates that GCMs capture the correct relationship between surface temperature and EIS in IV, but underestimate the trend because of their SST pattern.

The trend attribution from the GCM-based regression model confirm that the observed increase in $\langle{\rm EIS}\rangle_{\rm SEP}$ is dominated by remote warming. Nevertheless, the attribution map indicates different remote regions as the most important contributions to the trend (SI~Fig.~S7 shows the attribution map using the GCM-based regression model). Specifically, the North Pacific region that is highlighted by ERA5 (Fig.~\ref{5}C), is not reproduced by GCMs (SI~Fig.~S7E). This might reflect a spurious correlation in the limited ERA training data. The same region is highlighted when training on IV only (SI~Fig.~S7C), indicating that this correlation is not due to the large surface temperature trend in the North Pacific. Still, the contribution of local versus remote temperature changes is robust to choices in training data. Moreover, using Green's function sensitivity maps instead of regression-based sensitivity maps confirms that the observed trend in $\langle{\rm EIS}\rangle_{\rm SEP}$ is dominated by remote warming (SI~Fig.~S7G-H).

Even though the regression models were trained on data from the coupled Earth system, we do not quantify what processes are responsible for driving the observed surface temperature changes. For example, local surface cooling in the SEP could have been driven by remote effects \citep[e.g.,][]{Dong22,Kang23,Watanabe24,Zheng25}. Therefore, our results do not exclude remote influences on $\langle{\rm EIS}\rangle_{\rm region}$ by modulating local surface temperature patterns. The attribution maps quantify remote versus local contributions given the observed surface temperature pattern. 

Other regions of climatological marine low clouds show similar results (Table~\ref{tab1}), although recent trends are not as robust across different products (see SI~Fig.~S10 for 1980-2024 trends in different regions). The NEP has observed a significant increase in $\langle{\rm EIS}\rangle_{\rm NEP}$ since 1980, which was dominated by remote warming. In fact, the local warming in the NEP contributed negatively to the observed trend in $\langle{\rm EIS}\rangle_{\rm NEP}$, which was overcome by remote warming (SI~Fig.~S8 shows the attribution for the NEP). In the SEA, local and remote warming balanced each other, leading to a nonsignificant trend in $\langle{\rm EIS}\rangle_{\rm SEA}$, although our regression model predicts a slightly positive trend (Table~\ref{tab1}; see also SI~Fig.~S9 for the attribution in the SEA). This indicates that, even though the local contribution to the trend in $\langle{\rm EIS}\rangle_{\rm SEP}$ is not significant in 1980-2024, if the surface cooling trend in the SEP would turn into a warming trend in the future, the stabilizing trend in $\langle{\rm EIS}\rangle_{\rm SEP}$ would be reduced, all else equal. 

\section{Nonlinearity and implications for the pattern effect}

Recent work has indicated that the response of tropical deep convection to surface temperature changes is nonlinear in temperature and space \citep{Williams23,Bloch-Johnson24}. The same mechanisms might be at play here as well, because local $\langle{\rm EIS}\rangle$ is set to a large -- although not necessarily dominant -- degree by remote locations. We use realistic surface temperature patterns to train our regression model, as compared to localized patch perturbations \citep[as used in][]{Williams23,Bloch-Johnson24}. Therefore, our model implicitly includes nonlinearities due to the non-additivity of distinct localized perturbations. We quantify the degree to which the problem might be nonlinear by repeating the regression analyses with a convolutional neural network (CNN) instead of linear regression (see SI~Fig.~S11 for the CNN sensitivity maps). A CNN is a nonlinear regression model that has successfully been used to predict global radiation from spatial maps of surface temperature \citep[\ref{sec:app},][]{Rugenstein25,VanLoon25}. Because a CNN needs copious training data, we only apply it to GCMs. The CNN does not significantly improve the prediction of $\langle{\rm EIS}\rangle$ compared to linear regression and the sensitivity maps highlight the same regions. We conclude that nonlinear effects on $\langle{\rm EIS}\rangle$ are weak in the historical period, but cannot exclude that they may become more important in the future, when surface temperature anomalies are larger.

Our results give deeper insight into the mechanisms underlying the pattern effect. We quantify the relevance of remote locations onto $\langle{\rm EIS}\rangle$ and confirm that remote surface temperature changes modulate stability in the SEP more efficiently than local changes. However, the backbone effect of heating of the West Pacific Warm Pool on SEP stability is not the only mechanism at play. On a regional level, $\langle{\rm EIS}\rangle$ can be modified by surface temperature from a range of regions. This implies that locations with shallower or less frequent convection change $T_{700}$ as efficiently as deep convection in the West Pacific Warm Pool. This is in line with the ``circus tent'' model of the tropical atmosphere \citep{Williams23}, where strong temperature perturbations in moderately stable regions can increase tropical stability once the convective threshold is reached.

\section{Conclusion \& Outlook}

Regularized regression allows us to understand regional EIS. We move the analysis and understanding from near-global \citep[e.g.,][]{Myers23,Ceppi19} to local, in the regions that matter most for low cloud feedbacks and the pattern effect (Fig.~\ref{2}). Further work is necessary to quantify the influence of EIS versus other cloud controlling factors (most importantly surface temperature) onto the low clouds, their transition from stratus to cumulus, and their radiative effects. This might induce additional nonlinearities, because the effect of EIS on the shortwave cloud radiative effect might not be linear and the breakup of the stable boundary layer and cloud field likely follows a threshold behavior \citep{Wood12}. By studying these effects locally, we can gain process understanding of similarities and specifics of marine low cloud decks in different ocean basins.

Across ocean basins, regional EIS is similarly sensitive to local and remote surface temperature changes. Locally, a uniform surface warming of $\SI{1}{K}$ decreases $\langle{\rm EIS}\rangle_{\rm region}$ by $\sim\SI{-0.8}{K}$ in GCMs or $\sim\SI{-0.5}{K}$ to $\sim\SI{-0.6}{K}$ in reanalyses (summarized in SI~Table~S1). This indicates that local surface warming also warms the local free troposphere, destabilizing the atmosphere less than expected if the free troposphere would remain fixed. In contrast, a uniform remote warming of $\SI{1}{K}$ increases $\langle{\rm EIS}\rangle_{\rm region}$ by $\sim\SI{0.9}{K}$ both in GCMs and reanalyses (SI~Table~S1), indicating that remote surface warming is very efficient at changing the free tropospheric temperature in regions of climatological low cloud decks. Increased stability from remote warming is not dominated by the West Pacific Warm Pool as expected from Green's function experiments \citep[e.g.,][]{Zhou17,Dong19,Alessi23}, but other tropical and subtropical regions also matter (Fig.~\ref{2}). Because the sensitivity to remote surface temperature changes is larger than the sensitivity to local changes in reanalyses, global uniform warming increases stability in the SEP, NEP, and SEA. GCMs agree with ERA5 on this, except for the NEP, where local effects dominate (SI~Table~S1).

The sensitivities of EIS to surface temperature are robust across GCMs and reanalyses. The sensitivity maps trained on GCMs (with enough training data) and ERA5 (with limited data) are qualitatively similar (Fig.~\ref{2}). These sensitivity maps do not depend on GCMs simulating the correct surface temperature patterns, since we train them on IV only. Moreover, the statistical sensitivity maps highlight the same regions as model-based Green's function simulations (SI~Fig.~S6), indicating that they are picking up on the physical controls of surface temperature on EIS. GCMs and ERA5 agree on the remote sensitivity of $\langle{\rm EIS}\rangle_{\rm region}$ to surface warming (SI~Table~S1). This is remarkable, because convection is strongly parameterized in GCMs, which triggers tropical gravity waves and Rossby waves that transport heat in the free troposphere. The local sensitivity is stronger in GCMs than in ERA5, indicating that the surface is less strongly coupled to the free troposphere in GCMs than in ERA5. This might play a role in both the GCMs' large climatological mean-state biases and the spread in the cloud response to warming \citep{Ceppi24,Myers21,Myers23}.

We propose that tropical-to-extratropical teleconnections are necessary to explain the pattern effect and should be quantitatively assessed (Fig.~\ref{4}). The sensitivity maps of $\langle{\rm EIS}\rangle$ (Fig.~\ref{2}), together with localized warming experiments (Fig.~\ref{3}), confirm that tropical-to-extratropical teleconnections influence the free tropospheric temperature in regions of climatological marine low cloud decks \citep{Andrews18}. Rossby wave trains can counteract or add to the tropical free tropospheric warming from fast gravity waves (as in the WTGA), modulating stability in the subtropical East Pacific (see Fig.~\ref{4} for a schematic). This explains the seemingly weak sensitivity of $\langle{\rm EIS}\rangle_{\rm region}$ to surface warming in the West Pacific Warm Pool (Fig.~\ref{2}), because the resulting Rossby wave can have an opposite effect on $T_{700}$ than the WTGA (Fig.~\ref{3}B). The similarity between statistical sensitivity maps (Fig.~\ref{3}, which also include coupled processes) and Green's function sensitivity maps (SI~Fig.~S6) indicates that atmospheric processes are an important contributor to observed sensitivities. However, our regression framework cannot disentangle the relative importance or timescales of different processes, such as oceanic pathways \citep[e.g.,][]{Kang23,Zheng25}, land-sea contrast, or atmospheric teleconnections \citep[e.g.,][]{Zhou17,Andrews18,Dong19}. More work is needed to understand and quantify these different processes and their role in the pattern effect on low cloud feedbacks.

Finally, we attribute the observed increase in $\langle{\rm EIS}\rangle_{\rm SEP}$ since 1980. \cite{Myers23} argue that this trend has switched the global shortwave cloud radiative effect from negative to positive in recent decades. We find that local cooling only minimally contributed to the trend, which is instead dominated by remote warming (Fig.~\ref{5} and Table~\ref{tab1}). Our regression model trained on IV in GCMs attributes the trend correctly if we apply it to observed surface temperature patterns. This quantifies an unaddressed implication of the historical surface temperature bias in GCMs: coupled GCMs underestimate the trend in $\langle{\rm EIS}\rangle_{\rm SEP}$ because they do not simulate the correct surface temperature pattern, even though they capture the IV relationship between surface temperature and EIS correctly.

Our results lead to a better understanding of the drivers of marine low cloud feedbacks. Our regression method could help constrain low cloud feedbacks by training on GCM data and applying it to observed or plausible future surface temperature patterns, to overcome erroneous surface temperature trends simulated by GCMs \citep[e.g.,][]{Seager22,Wills22}. In the near future, the observed SEP warming will contribute to a decrease in local EIS. However, whether and when the EIS trend will reverse will more likely be set by the exact pattern of remote warming. This subtle interplay between local and remote effects will determine the efficiency of the low cloud decks to cool the Earth.

%
%
\appendix

\section{Regression and attribution details} \label{sec:app}

\subsection{Linear ridge regression}

For climate model data, we use Tensorflow \citep{Tensorflow} to perform ridge regression, using a ridge parameter $\alpha = 0.25$, a learning parameter of $5\times10^{-6}$, and a batch size of $32$. We use detrended data (calculated by removing the ensemble mean) in all years 1921-2014 to train, using 24 ensemble members for training, 3 for validation, and 3 for testing. The least-squares loss function is optimized using the Adam optimizer \citep{Kingma17} and an early stopping criterion from the least-squares loss in the validation dataset, to further reduce the risk of overfitting. We train a linear regression model on data from all four climate models simultaneously and on each model separately, using the same training parameters. When training on all models simultaneously, we use the same amount of ensemble members and years from each model (i.e., $4\times 24$ members for training, $4\times 3$ for validation, and $4\times 3$ for testing). 

For ERA5 data, we use scikit-learn \citep{scikit-learn} to perform ridge regression with leave-one-out cross-validation to determine the optimal value for the ridge parameter. We train two different regression models: one on high-pass filtered ERA5 data (i.e., high-frequency variability) and one on unfiltered ERA5 data (i.e., including low-frequency variability). For unfiltered ERA5 data, of all 85 years in 1940-2024, we randomly select 15 years for testing, and use 70 years for training and validation (see SI~Fig.~S12 for years used). For the filtered data, we only have 77 years, but still select 15 years for testing, leaving 62 years for training and validation.

\subsection{Convolutional neural network}

In addition to linear regression, we train a convolutional neural network (CNN) to predict $\langle{\rm EIS}\rangle$ from $T$. The CNN is nonlinear and is designed to recognize spatial patterns in the data. We use the same architecture as \cite{VanLoon25}, which consists of two convolutional layers and two fully connected layers. We find no significant improvement with other architectures. See \cite{Rugenstein25} and \cite{VanLoon25} for more details.

\subsection{Trend analysis}

We use ordinary least squares (OLS) regression to determine the trend in $\langle{\rm EIS}\rangle$ in 1980-2024. We do this for three reanalysis datasets and all available ensemble members of four GCMs, as described in section~\ref{methods}. For reanalyses, the standard errors of the predicted slopes are converted to a 5-95\% confidence interval by assuming a $t$-distribution with 43 degrees of freedom. For the GCMs, the 5-95\% confidence interval is calculated from percentiles of the ensemble distribution.

Predicted trends are calculated by first applying the regression model to surface temperature data in 1980-2024 and then computing the trends in the predicted $\langle{\rm EIS}\rangle$ timeseries. For a given regression model (unfiltered ERA5, filtered ERA5, or GCM internal variability), this is done with $T$ from each reanalysis and gridded SST product separately.

In Fig.~\ref{5}D, we only show the OLS slopes, not the confidence intervals. That is, the bars span from the minimum to maximum predicted slope, disregarding the standard error associated with the OLS regression. The reanalysis bar (green) contains only three predictions (from MERRA2, JRA-3Q, and ERA5, indicated by crosses). The GCM bar contains 240 predictions, one for each ensemble member, spanning four different models and two different future scenarios. The regression bar contains all predictions from the three different regression models applied to six estimates of historical $T$, totaling 18 predictions. See SI~Fig.~S10 for a breakdown of trends into different models, scenarios, and $T$ estimates.

\subsection{Attribution}

We use the ``input times gradient'' method to attribute the regions that contribute to the trend in $\langle{\rm EIS}\rangle$ \citep{Shrikumar17, Mamalakis22}. First, we calculate the local trend in $T$ using ordinary least squares regression. Then, we multiply the local trend in $T$ with the sensitivity map ${\rm d}\langle{\rm EIS}\rangle/{\rm d}T$ to obtain the trend attribution map. 

We also calculate the attribution maps for every year separately, by multiplying the sensitivity map ${\rm d}\langle{\rm EIS}\rangle/{\rm d}T$ with the local temperature anomaly in each year. Summing these attribution maps over all ocean areas results in the predicted $\langle{\rm EIS}\rangle$ timeseries from the regression model. By only summing over a certain region (e.g., the SEP in Fig.~\ref{5}), we can calculate the contribution of that region to the predicted timeseries (blue line in Fig.~\ref{5}D).

%
%

\section*{Open Research Section}
All climate model data is standard CMIP model output, and is made freely available by the Earth System Grid Federation (ESGF) at https://esgf-node.llnl.gov/. Reanalysis data is available from \cite{Hersbach23}, \cite{cisl_rda_dsd640002}, and \cite{merra2data}; gridded SST products from \cite{COBE2}, \cite{NOAAGlobalTemp}, and \cite{cisl_rda_dsd277003}. All reported results were analyzed using Python-3.10. All code will be made available at the time of acceptance of the manuscript.

\section*{Conflict of Interest disclosure}
The authors declare there are no conflicts of interest for this manuscript.

\acknowledgments
We thank Elizabeth Barnes, Paulo Ceppi, Leif Fredericks, Timothy Myers, Cristian Proistosescu, and Levi Silvers for discussions; Benjamin O. Johnson, Juliet Pilewskie, David Randall, and David W. J. Thompson for thoughts on an initial version of the manuscript; and Madeline Mendell for help with Fig.~\ref{1}C. This work was supported, in part, by the Regional and Global Model Analysis program area of the U.S. Department of Energy's Office of Biological and Environmental Research as part of the Program for Coordinated Model Diagnosis and Intercomparison.

\clearpage
\section*{Supporting Information}
\setcounter{figure}{0}
\setcounter{table}{0}
\renewcommand{\figurename}{{\bf SI Fig.}}
\renewcommand{\tablename}{{\bf SI Table}}
\renewcommand{\thefigure}{{\bf S\arabic{figure}}}
\renewcommand{\thetable}{{\bf S\arabic{table}}}

\begin{figure*}[ht]
    \centering
    \includegraphics{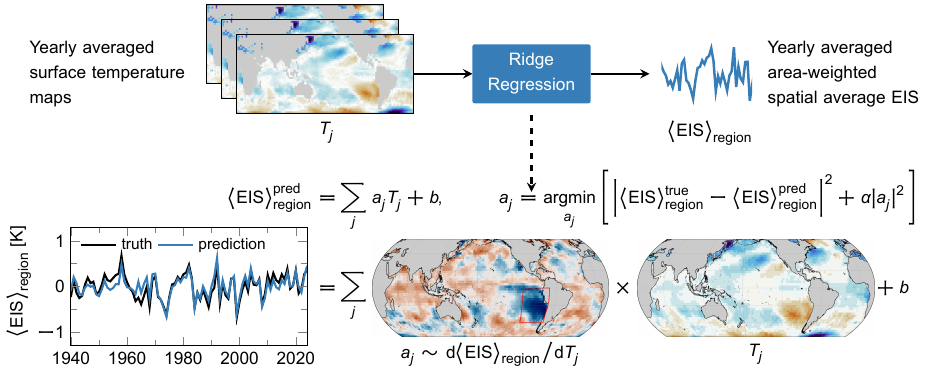}
    \caption{Schematic of the ridge regression method to determine the sensitivity of estimated inversion strength (EIS) to surface temperature ($T$). The coefficients of the linear regression model are interpreted as the sensitivity map $a_j \sim {\rm d}\langle{\rm EIS}\rangle_{\rm region}/{\rm d}T_j$. The vertical delimiters $\vert .\vert$ indicate an $L_2$ norm and $\alpha$ is the ridge parameter.}
    \label{SIfig:RidgeRegression}
\end{figure*}

\begin{figure*}[ht]
    \centering
    \includegraphics{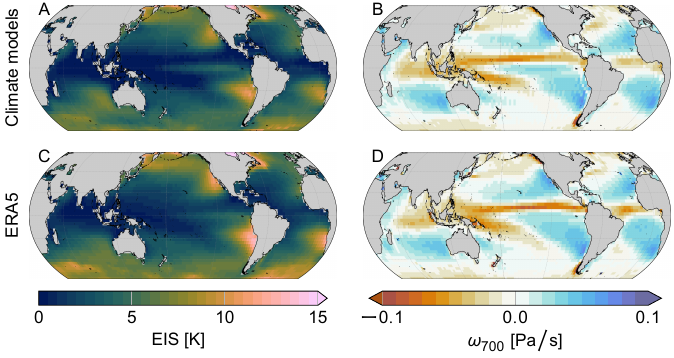}
    \caption{Climatology of estimated inversion strength (EIS) and vertical velocity at $\SI{700}{hPa}$ ($\omega_{700}$) in 1991-2014. Top row shows the climatology averaged over three coupled climate models (CanESM5, MIROC6, and MPI-ESM1.2-LR). Bottom row shows the climatology in ERA5.}
    \label{SIfig:Climatology}
\end{figure*}

\begin{figure*}[ht]
    \centering
    \includegraphics{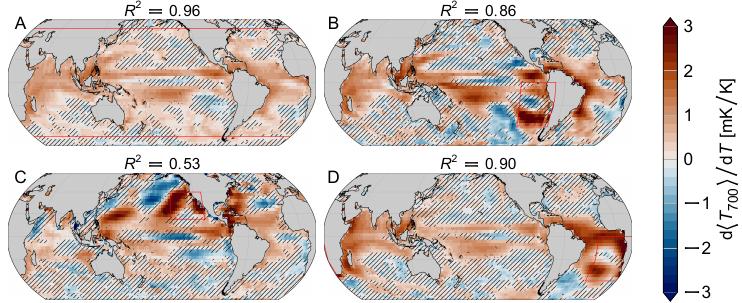}
    \caption{Controls of near-global and regional air temperature at $\SI{700}{hPa}$. 
    Sensitivity to surface temperature of $\langle T_{700} \rangle$ averaged over regions indicated by red boxes: (A) near global, (B) Southeast Pacific, (C) Northeast Pacific, and (D) Southeast Atlantic, based on ridge regression on data from four climate models. Hatching indicates regions where sensitivity maps obtained from the four models separately do not agree on the sign.}
    \label{SIfig:Sensitivity_T700}
\end{figure*}

\begin{figure*}[ht]
    \centering
    \includegraphics{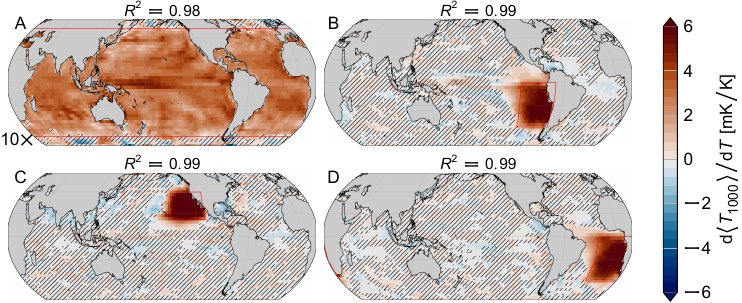}
    \caption{Controls of near-global and regional air temperature at $\SI{1000}{hPa}$. 
    Sensitivity to surface temperature of $\langle T_{1000} \rangle$ averaged over regions indicated by red boxes: (A) near global, (B) Southeast Pacific, (C) Northeast Pacific, and (D) Southeast Atlantic, based on ridge regression on data from four climate models. Note that the near global map (A) is multiplied by a factor of 10 to use the same color bar. Hatching indicates regions where sensitivity maps obtained from the four models separately do not agree on the sign.}
    \label{SIfig:Sensitivity_T1000}
\end{figure*}

\begin{figure*}[ht]
    \centering
    \includegraphics[width=\textwidth]{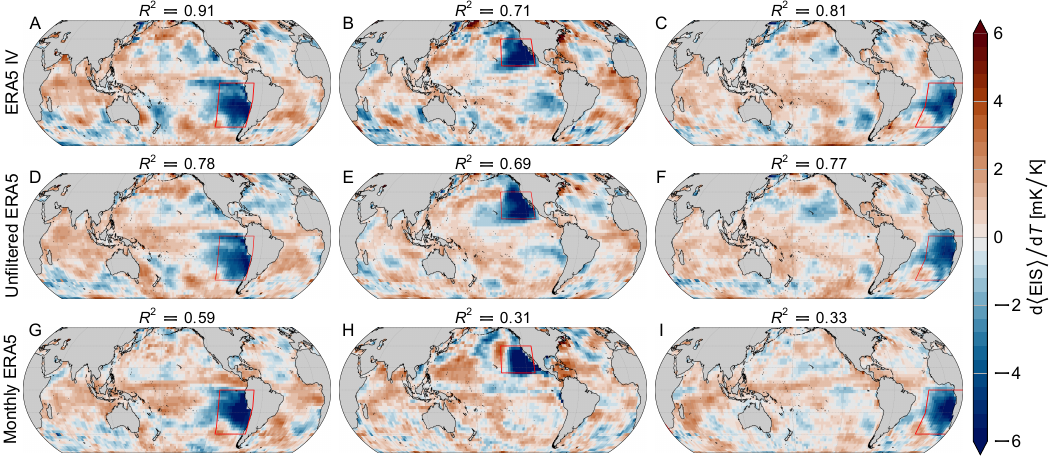}
    \caption{Comparison of sensitivity of regional estimated inversion strength to surface temperature in annual internal variability (IV, estimated by filtering the data with a 10-year high-pass filter), annual unfiltered data, and monthly data. 
    Top row shows the sensitivity to surface temperature of $\langle {\rm EIS} \rangle$ averaged over regions indicated by red boxes: (A) Southeast Pacific, (B) Northeast Pacific, (C) Southeast Atlantic, based on ridge regression on data from ERA5 that was detrended by applying a high-pass filter with a cutoff of 10 years. 
    Middle row shows the sensitivity maps in the same regions, but using unfiltered ERA5 data. 
    Bottom row uses unfiltered ERA5 data on monthly timescales, instead of annual averages. 
    $R^2$ values are calculated for held-back testing members and displayed above each map.
    }
    \label{SIfig:RegionalSensitivity_yearly_ERA5_IVvsTR}
\end{figure*}

\begin{figure*}[ht]
    \centering
    \includegraphics{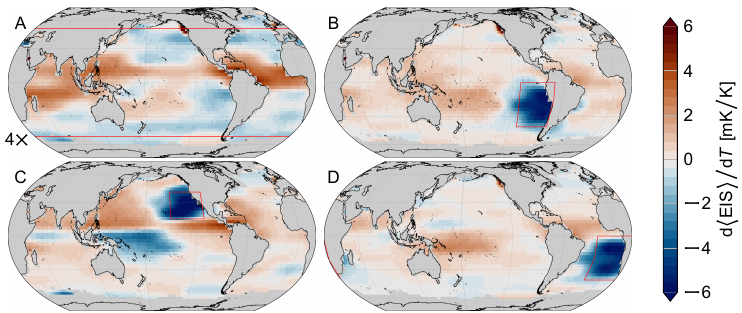}
    \caption{Sensitivity maps from Green's function experiments in ECHAM6, for different regions discussed in the main text. Note that the near global map (A) is multiplied by a factor of 4 to use the same color bar.}
    \label{SIfig:GFs}
\end{figure*}

\begin{figure*}[ht]
    \centering
    \includegraphics{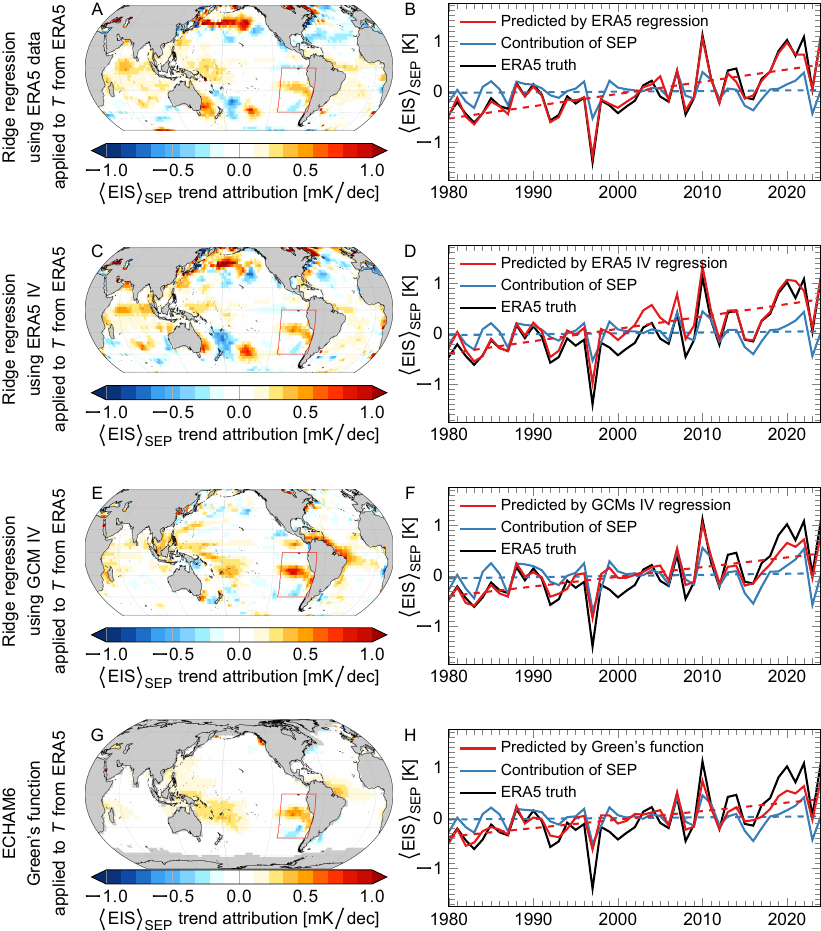}
    \caption{Attribution of the estimated inversion strength trend in 1980-2024 in the Southeast Pacific. 
    Left plots show attribution maps of $\langle{\rm EIS}\rangle_{\rm SEP}$, EIS averaged over the red box in the Southeast Pacific (SEP), calculated by multiplying the observed surface temperature trend (main text, Fig.~4A) with the sensitivity map from (A) ridge regression on ERA5 data (main text, Fig.~4B); (B) ridge regression on 10-year high-pass-filtered ERA5 data (main text, Fig.~2D); (C) internal variability in four climate models (main text, Fig.~2A); and (D) ECHAM6 Green's function (Fig.~\ref{SIfig:GFs}B). Right plots show the observed $\langle{\rm EIS}\rangle_{\rm SEP}$ in 1980-2014 (black), compared to the prediction from the sensitivity maps (red). Blue line shows the contribution from local temperature changes only, calculated by multiplying the observed surface temperature with the sensitivity map and summing over the area indicated by the red box in panel A.
    All values are anomalies with respect to the 1980-2024 average. Dashed lines indicate the trend in 1980-2024.}
    \label{SIfig:Attribution_ERA5_SEP_Model_vs_ERA5}
\end{figure*}

\begin{figure*}[ht]
    \centering
    \includegraphics{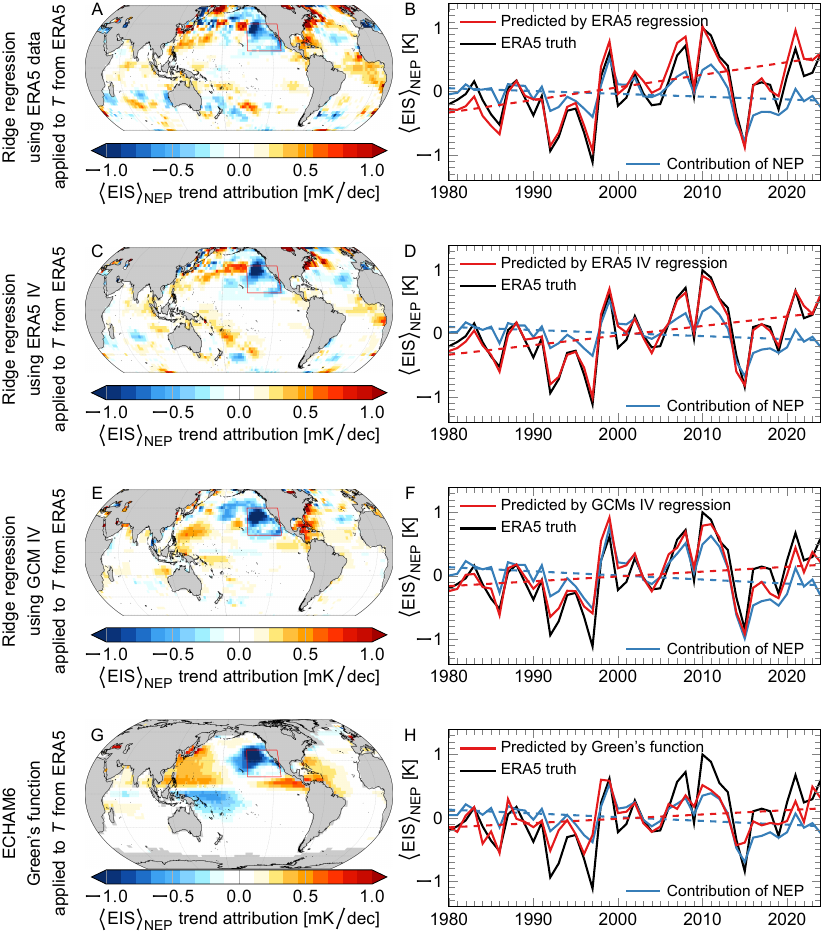}
    \caption{Same as Fig.~\ref{SIfig:Attribution_ERA5_SEP_Model_vs_ERA5}, but for the Northeast Pacific.}
    \label{SIfig:Attribution_ERA5_NEP_Model_vs_ERA5}
\end{figure*}

\begin{figure*}[ht]
    \centering
    \includegraphics{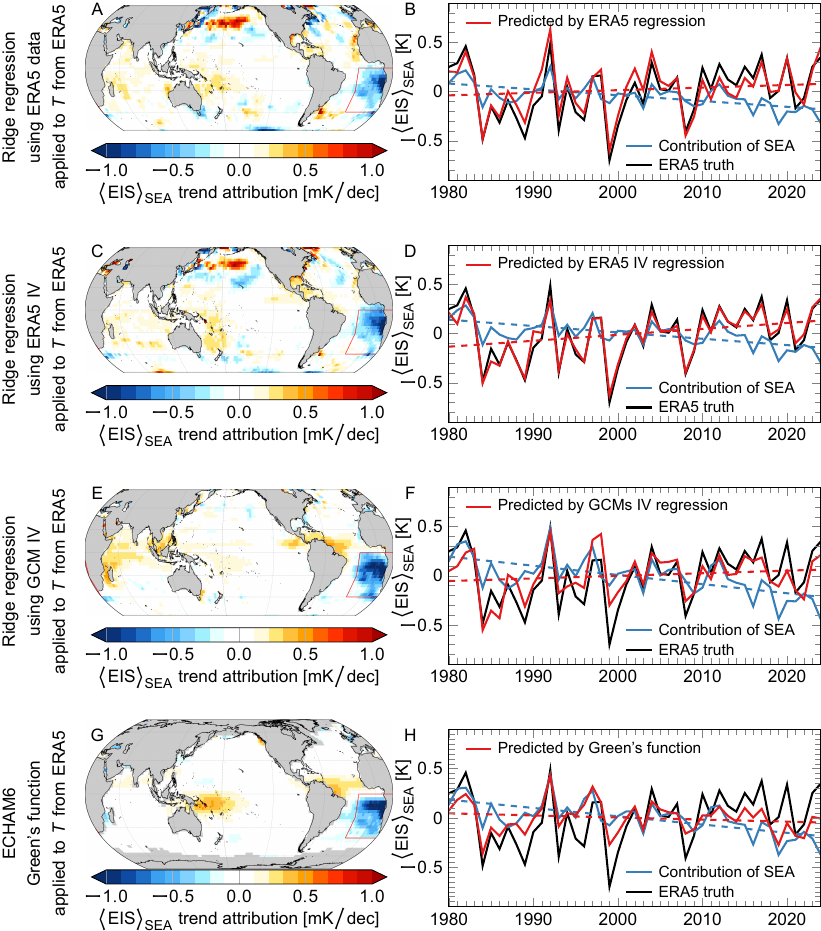}
    \caption{Same as Fig.~\ref{SIfig:Attribution_ERA5_SEP_Model_vs_ERA5}, but for the Southeast Atlantic.}
    \label{SIfig:Attribution_ERA5_SEA_Model_vs_ERA5}
\end{figure*}

\begin{figure*}[ht]
    \centering
    \includegraphics[width=\textwidth]{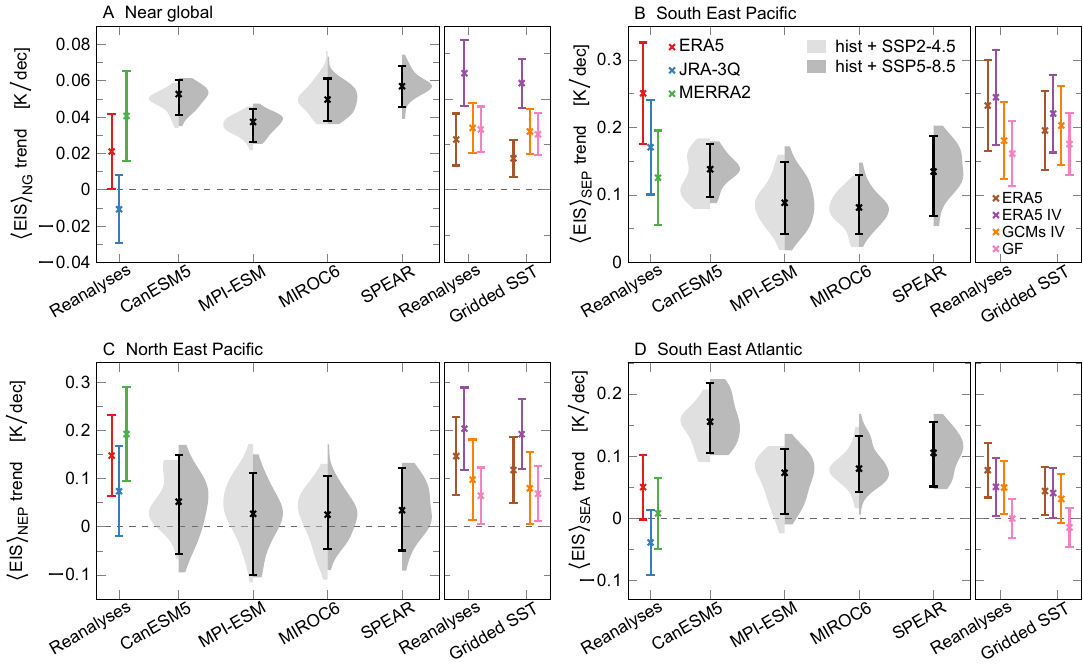}
    \caption{Estimated inversion strength (EIS) trends in 1980-2024 in different regions of interest: (A) near global, (B) Southeast Pacific, (C) Northeast Pacific, and (D) Southeast Atlantic. In each panel, the left plot shows the trends based on three reanalysis products and four GCMs. Shading shows the distribution of trends in different ensemble members. Right plot shows the predicted trends by applying the regression model trained on ERA5 data including the trend (brown), ERA5 internal variability (purple) and GCMs internal variability (orange) to different surface temperature datasets. Pink shows the ECHAM6 Green's function (GF) applied to surface temperature datasets. The cross indicates the results averaged over three reanalyses (ERA5, JRA-3Q, and MERRA-2) and averaged over three gridded SST datasets (COBE2, NOAAGlobalTemp, and HadISST), while the error bars indicate the $\SI{5}{\percent}$-$\SI{95}{\percent}$ confidence bounds.}
    \label{SIfig:EIS_trends_pred}
\end{figure*}

\begin{figure*}[ht]
    \centering
    \includegraphics[width=\textwidth]{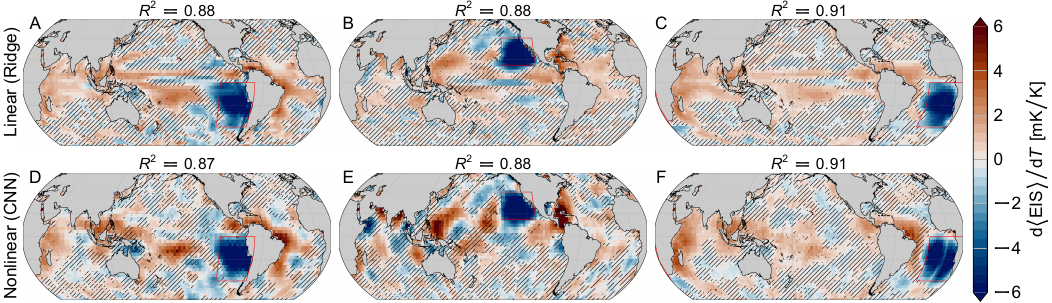}
    \caption{Nonlinearity of the sensitivity of estimated inversion strength. 
    Top row shows the same maps as Fig.~2A-C in the main text: the sensitivity to surface temperature of $\langle {\rm EIS} \rangle$ averaged over regions indicated by red boxes: (A) Southeast Pacific, (B) Northeast Pacific, (C) Southeast Atlantic, based on ridge regression on data from four climate models.
    Bottom row shows the same maps, but using a nonlinear convolutional neural network (CNN) to predict $\langle {\rm EIS} \rangle$ from $T$. The CNN is trained on the same data as the ridge regression. 
    Hatching indicates regions where sensitivity maps obtained from the four models separately do not agree on the sign. $R^2$ values are calculated for held-back testing members and displayed above each map.} 
    \label{SIfig:RegionalSensitivity_CNN_vs_Linear}
\end{figure*}

\begin{figure*}[ht]
    \centering
    \includegraphics[width=\textwidth]{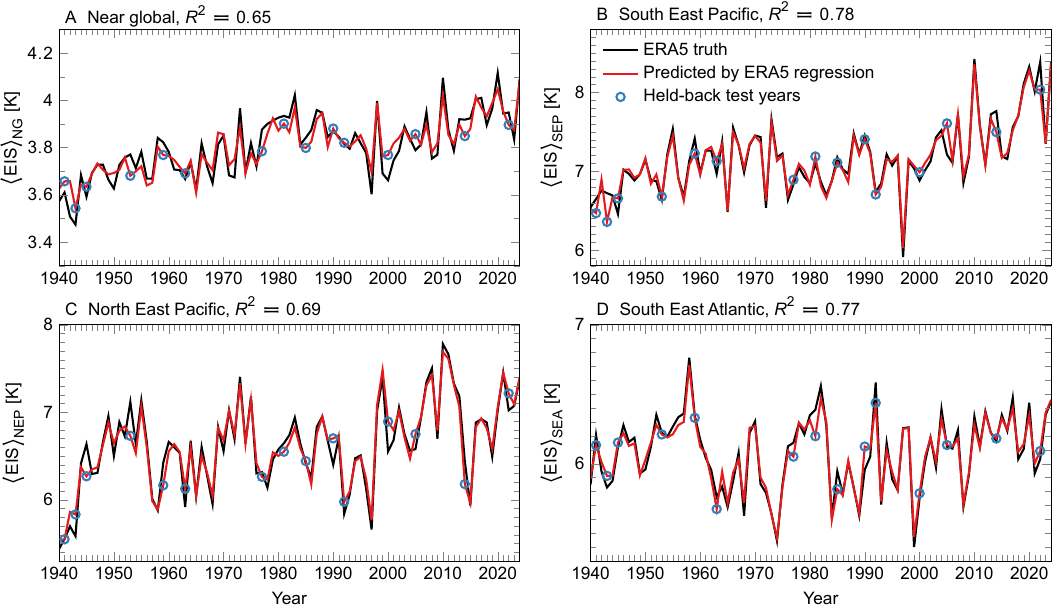}
    \caption{Regional and near-global EIS in ERA5. Black lines show the average EIS from ERA5 in the (A) near global, (B) Southeast Pacific, (C) Northeast Pacific, and (D) Southeast Atlantic. The red line shows the prediction from ridge regression on ERA5 data. Blue circles indicate the years that were held back for testing, and were thus not used to train the regression model. $R^2$ values are calculated with testing members only.}
    \label{SIfig:EIS_ERA5_testmems}
\end{figure*}

\clearpage
\begin{table*}[ht]
    \centering
    \caption{Local versus remote contribution to the sensitivity of $\langle{\rm EIS}\rangle$ to uniform warming. Calculated by summing the sensitivity map of the three different regression models (ERA5 with trend, high-pass filtered ERA5, and internal variability in GCMs) over all grid boxes (total), over the local region (local), and over all remote regions (remote).} 
    \begin{tabular}[h]{c c | c c c}
        Region & Regression model & Total & Local & Remote \\\hline
        NG & ERA5 & $0.26$ &   &   \\
        NG & ERA5 IV & $0.31$ &   &   \\
        NG & GCM IV & $0.11$ &   &   \\\hline
        SEP & ERA5 & $0.54$ & $-0.58$ & $1.1$ \\
        SEP & ERA5 IV & $0.33$ & $-0.62$ & $0.94$ \\
        SEP & GCM IV & $0.13$ & $-0.81$ & $0.94$ \\\hline
        NEP & ERA5 & $0.39$ & $-0.55$ & $0.94$ \\
        NEP & ERA5 IV & $0.33$ & $-0.58$ & $0.91$ \\
        NEP & GCM IV & $-0.13$ & $-0.82$ & $0.69$ \\\hline
        SEA & ERA5 & $0.49$ & $-0.46$ & $0.95$ \\
        SEA & ERA5 IV & $0.47$ & $-0.43$ & $0.90$ \\
        SEA & GCM IV & $0.14$ & $-0.76$ & $0.90$
    \end{tabular}
    \label{SItab1}
\end{table*}

\begin{table*}[ht]
    \centering
    \caption{Datasets used in this study.} 
    \begin{tabular}[h]{r | c c c c c}
        Name & Scenario & Period & \#Members & Resolution & Reference \\\hline
        CanESM5 & Historical & 1850-2014 & 40 & 2.8°x2.8° & \citep{CanESM5} \\
         & SSP2-4.5 & 2015-2100 & 25 & 2.8°x2.8° &  \\
         & SSP5-8.5 & 2015-2100 & 25 & 2.8°x2.8° &  \\
        GFDL-SPEAR-MED & Historical & 1921-2014 & 30 & 0.625°x0.5° & \citep{GFDLSPEARMED} \\
         & SSP5-8.5 & 2015-2100 & 30 & 0.625°x0.5° &  \\
        MIROC6 & Historical & 1850-2014 & 50 & 1.4°x1.4° & \citep{MIROC6} \\
         & SSP2-4.5 & 2015-2100 & 50 & 1.4°x1.4° &  \\
         & SSP5-8.5 & 2015-2100 & 50 & 1.4°x1.4° &  \\
        MPI-ESM1.2-LR & Historical & 1850-2014 & 50 & 1.9°x1.8° & \citep{MPI12} \\
         & SSP2-4.5 & 2015-2100 & 30 & 1.9°x1.8° &  \\
         & SSP5-8.5 & 2015-2100 & 30 & 1.9°x1.8° &  \\\hline
        ERA5 & Historical & 1940-2024 & 1 & 0.25°x0.25° & \citep{Hersbach20} \\
        JRA-3Q & Historical & 1948-2024 & 1 & 0.375°x0.375° & \citep{Kosaka24} \\
        MERRA2 & Historical & 1980-2024 & 1 & 0.625°x0.5° & \citep{Gelaro2017} \\\hline
        COBE2 & Historical & 1850-2024 & 1 & 1°x1° & \citep{COBE2} \\
        NOAAGlobalTemp & Historical & 1850-2024 & 1 & 5°x5° & \citep{NOAAGlobalTemp} \\
        HadISST & Historical & 1870-2024 & 1 & 1°x1° & \citep{Rayner03}
    \end{tabular}
    \label{SItab2}
\end{table*}

\clearpage
\setstretch{1.1}
\bibliographystyle{ametsocV6}

\end{document}